\documentclass[aps,prl,twocolumn,showpacs,superscriptaddress,groupedaddress]{revtex4}  % for review and submission
\usepackage{graphicx}  % needed for figures
\usepackage{dcolumn}   % needed for some tables
\usepackage{bm}        % for math
\usepackage{amssymb, amsmath}   % for math
\usepackage{epsfig}
\usepackage{subfigure}

\begin{document}

% The following information is for internal review, please remove them for submission
%\leftline{Primary authors: XXX}
%\leftline{To be submitted to (PRL, PRD-RC, PRD, PLB; choose one.)}
%\leftline{Comment to {\tt d0-run2eb-nnn@fnal.gov} by xxx, yyy}
%\centerline{\em D\O\ INTERNAL DOCUMENT -- NOT FOR PUBLIC DISTRIBUTION}

% the following line is for submission, including submission to the arXiv!!
%\hspace{5.2in} \mbox{Fermilab-Pub-04/xxx-E}
\author{Yan-Jiun Chen}
\affiliation{Department of Physics, Cornell University, Ithaca NY 14850}
\author{Natalie M. Paquette}
\affiliation{SITP, Department of Physics, Stanford University, Stanford CA 94305}
\author{Benjamin B. Machta}
\affiliation{Lewis-Sigler Institute for Integrative Genomics, Princeton University, Princeton NJ 08544}
\author{James P. Sethna}
\affiliation{Department of Physics, Cornell University, Ithaca NY 14850}

%\title{2D Ising Model: Precise functional form for the spin-spin correlation function}
\title{Universal scaling function for the two-dimensional Ising model in an external field: A pragmatic approach}

\begin{abstract}
We report an effective functional form for the spin-spin correlation function of the 2D Ising model as a function of temperature and field. Although the Ising model has been well studied, no analytical result for the spin-spin correlation function exists for arbitrary magnetic fields and temperatures. We show the validity of our form by comparison with simulations using the Wolff algorithm, and obtain remarkable precision by including analytic corrections to scaling. Given recent interest in comparing biomembrane heterogeneity to Ising criticality, our spin-spin correlation function may be used as a predictive quantitative measure for FRET or NMR membrane experiments. 
\end{abstract}
%\pacs{87.15.kt, 87.15.Ya,87.16.dt}
\maketitle
%XXX changes to this par:  
%peculiar-> unique
The two-dimensional (2D) Ising model occupies a unique place in statistical physics. As the simplest example of a system displaying nontrivial critical phenomenon it has long been a testing ground for theoretical and computational methods, having spawned thousands of papers. In addition, it is the canonical member of the 2D Ising universality class. Members of the 2D Ising universality class include many current experimental systems, some ({e.g.} liquid-vapor phase transitions or membrane phase diagrams) far removed from its original conception as a simple model of ferromagnetism.

In the vicinity of its critical point the spin-spin correlation function for the 2D Ising model approaches a two parameter universal scaling function
\begin{equation}
C(r| H, T) = \langle \sigma_r \sigma_0 \rangle = r^{-\eta} {\cal C} (r/t^{-\nu}, h/t^{\beta \delta})
\label{eq:scalingCorr}
\end{equation}
with $\eta=1/4$, $\nu=1$, $\beta = 1/8$, $\delta = 15$, reduced temperature $t=\frac{T-T_c}{T}$ and field $h = \beta H = H/T$.  The form of the universal function ${\cal C}$ is known along two lines through its parameter space:  when $H=0$ it can be written in terms of an integral over Painlev{\'e} transcendents~\cite{mccoy1973two} and when $T=T_c$ there exists a complete asymptotic expansion that uses exact results from integrable field theory~\cite{zamolodchikov1989integrals,delfino1995spin}.   Approximate functional forms for the correlation function and the correlation length have been developed for the 3D Ising model in a field in momentum space~\cite{tarko1975theory, fisher1967theory}.
% XXX YJ and I don't understand what this means? orrick doesn't discuss a field,
% and the abstract for zamo 89 doesn't mention anything except T=Tc. 
%, and field theoretic and transfer matrix techniques have been used to extend 2D results in a field~\cite{zamolodchikov1989integrals, orrick2001critical}. 
Here we leverage the known exact results and a high precision approximate form for the free energy~\cite{caselle2001critical} to develop an elegant interpolation for the scaling form for the 2D Ising correlation function in an external field.

In addition to filling a surprising gap in the theoretical Ising literature, our results are of practical relevance for the interpretation of experiments in multicomponent lipid membranes. Phase diagrams for these membranes often contain miscibility critical points in the 2D Ising universality class~\cite{veatch2007critical,Honerkamp-Smith2009overview}.  Recent experiments suggest that cells maybe tuning their own membranes to the proximity of this critical point~\cite{Veatch2008Vivo} suggesting they may be taking advantage of criticality's unique physics~\cite{machta2011minimal,Machta2012Casimir}.  NMR~\cite{veatch2007critical}, FRET~\cite{Heberle2010comparison}, and fluorescence microscopy~\cite{Veatch2005Spots} all yield observables that are simply related to the underlying membrane's correlation functions.  Although scattering experiments in typical three dimensional systems more naturally probe the Fourier transform of $C(r)$, membrane probes more typically measure real-space properties.  With our scaling forms it will be possible to map the composition and temperature parameters of these membranes onto the Ising axis of $t$ and $M$ (the magnetization).

\begin{figure}
	\includegraphics[scale=0.35]{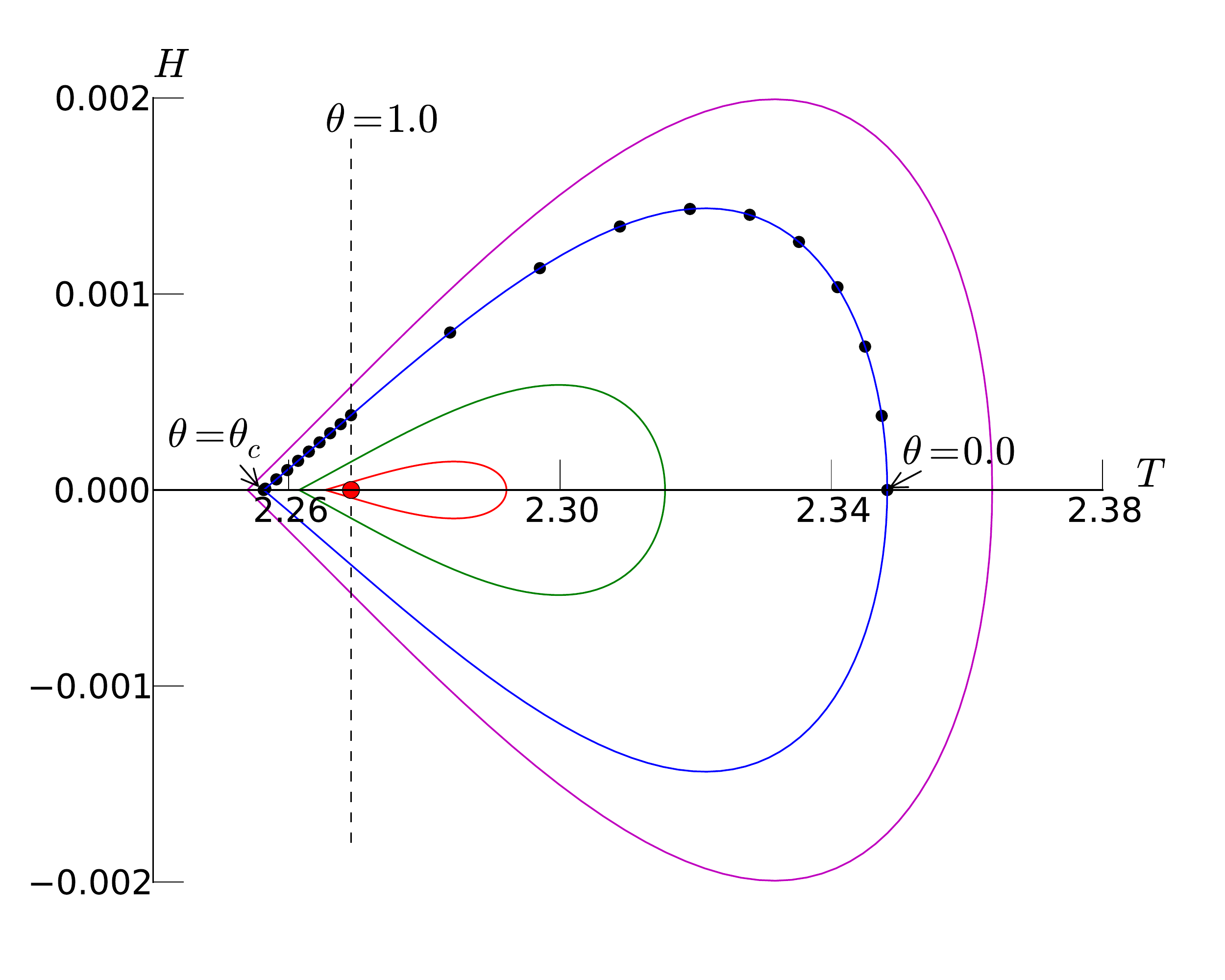}
	\caption{\label{fig:ConstantR}  Curves of constant `polar coordinate' $R$ are plotted against the Ising parameters of $T$ and $H$, from outwards in: $R=0.4$ (magenta), $R=0.336737$ (blue), $R=0.2$ (green), $R=0.1$ (red). The critical point is labeled by the red circle.  We compare to Monte-Carlo results for $1024 \times 1024$ simulations with parameters given by the blue dots. Note that the half-plane $T<Tc$ spans a small range of $\theta$, namely from $\theta=1$ to $\theta=\theta_c \approx 1.08144$~\cite{caselle2001critical}}
\end{figure}

It is useful to describe the scaling behavior near the critical points using the Schofield `polar coordinate' parameterization~\cite{schofield1969parametric, schofield1969correlation, caselle2001critical}, which is expressed as follows:
\begin{equation}
\begin{split}
t &= \frac{T-T_c}{T} = R(1-\theta^{2}) \\
h &= H/T = h_{0} R^{\beta\delta}h(\theta) \\
M &= m_{0} R^{\beta} \theta
\end{split}
\end{equation}
Here, we use Caselle's~\cite{caselle2001critical} high precision form for $h(\theta)$:
\begin{align} 
h(\theta) &= ( \theta - \theta^3/1.16951) (1 - 0.222389 \theta^2 \cr
&- 0.043547\theta^4 - 0.014809\theta^6 - 0.007168\theta^8). 
\label{eq:htheta}
\end{align}
We take $m_0 = 0.90545$~\cite{YangMagnetization1952} and $h_0 = 0.940647$~\cite{caselle2001critical}. Figure~\ref{fig:ConstantR} shows a representation of the coordinate transform, with curves of constant $R$ plotted in $(T, H)$. Exact scaling results exist at three points along each curve, $\theta = 0$ ($H=0$, $T>T_c$), $\theta=\theta_c$ ($H=0$, $T < T_c$), $\theta=1.0$ ($T=T_c$). Other systems (like membrane miscibility phase diagrams) can be treated by suitable mappings of their control variables into $(R, \theta)$.

We would like to compare our functional form to simulation results from systems that are close enough to the critical point to be in the scaling regime yet which have a correlation length small compared to our lattice size $\mathrm L=1024$ so as to minimize finite-size corrections. Therefore we run simulations at a range of $\theta$ values at a fixed $R$ value chosen to have correlation lengths between $\xi_+  \approx \mathrm{L}/60$ and $\xi_- \approx \mathrm{L}/10$ above and below $T_c$.  Interpolating between known results along $h=0$ above and below $T=T_c$~\cite{wu1976spin}, and using a high-precision form~\cite{caselle2001critical} of the susceptibility as an integration constraint, we are able to arrive at an interpolating correlation function that matches our simulation data for all values of $\theta$. 

As one component of the functional form, we need an interpolating form for the asymptotic correlation length $\xi(H,T) = \xi(R, \theta)$ giving the long-distance exponential decay of the correlation function. Here we define  a scaling variable $s=r/\xi(H,T) $, where $s \approx (4/T_c) (r/ t^{-\nu}) $ along $H=0$ and $s \approx r (h/h_s)^{8/15}$ with $ h_s  = \left(\frac{\Gamma(2/3) \Gamma(8/15)}{ 4 \sin(\pi/5) \pi \Gamma(1/5)}\right)^{15/8} \left( \frac{\Gamma(1/4) \Gamma^2(3/16)}{4 \pi^2 \Gamma(3/4) \Gamma^2 (13/16)} \right)^{1/2} $ along $T=T_c$~\cite{delfino1995spin}.  We design an even polynomial in $\theta$ in the form of $\xi(R, \theta) = \Xi(\theta)/R$, since $\xi \sim t^{-\nu} \sim R^{-\nu}$ where $\nu=1$.
\begin{equation}
\Xi(\theta) = a_0 + a_1 \theta^2 + a_2 \theta^{34}
\label{eq:XiFunction}
\end{equation} 
Matching the two known values at $H=0$ and the value at $T=T_c$ we fix $a_0$, $a_1$, and $a_2$. $a_0 = 0.567296$, $a_1 =0.0284915$ and $a_2 = 0.19171$. The power $34$ is taken from a fit of $a_0 + a_1 \theta^2 + a_2 \theta^n$ to known values at $\Xi(0)$, $\Xi(1)$, $\Xi(\theta_c)$, where $34$ is the smallest even power that allows $\Xi(\theta)$ to be monotonic with increasing $\theta$. This large power is likely due to the strong asymmetry of the Schofield coordinates (Figure~\ref{fig:ConstantR}), which compress the range $T<T_c$ into $1 < \theta<\theta_c \approx 1.08144$. 

Now, with the scaling variable $r/\xi = rR/\Xi(\theta)$ we can further use known scaling solutions to find an interpolating form for the correlation function. We design a function that interpolates between the exact scaling solution~\cite{wu1976spin} $F_{+}(s)$ at $T>T_c$ and $F_{-}(s) - M^2$ for $T<T_c$, replacing the scaling variable with our form $s = rR/\Xi(\theta)$.  The function $f(\theta)$ controls the interpolation, and is designed such that $f(0)=0$ and $f(\theta_c)=1$.  Outside of the interpolation, we add back the magnetization in Schofield coordinates with $(m_0 R^\beta \theta)^2$.
%\begin{equation}
%\begin{array}{rl}
%C(r|R,\theta) = &r^{-1/4} \Bigg[ \left( 1-f(\theta) \right)F_{+} \left(\frac{r}{R \Xi(\theta)}\right)\\ 
%+ & f(\theta)\left(F_{-} \left(\frac{r}{R \Xi(\theta)}\right) - 2^{3/8} \left(\frac{r}{R \Xi(\theta)}\right)^{1/4} \right) \Bigg]\\
%+ & (m_0 R^{1/8} \theta)^2
%\end{array}
%\label{eq:interpolatingC}
%\end{equation}
%\begin{equation}
%\begin{array}{l}
%C(r|R,\theta) = r^{-1/4} \Bigg[ \left( 1-f(\theta) \right) F_{+} \left(\frac{r}{R \Xi(\theta)}\right)+\\ 
%f(\theta) \left(F_{-} \left(\frac{r}{R \Xi(\theta)}\right) - 2^{3/8} \left(\frac{r}{R \Xi(\theta)}\right)^{1/4} \right) \Bigg]+ (m_0 R^{1/8} \theta)^2
%\end{array}
%\label{eq:interpolatingC}
%\end{equation}

\begin{equation}
\begin{array}{l}
C(r|R,\theta) =(m_0 R^{1/8} \theta)^2+\\ r^{-1/4} \Big[ \left( 1-f(\theta) \right) F_{+} \left(s\right)+
f(\theta) \left(F_{-} \left(s\right) - 2^{3/8} s^{1/4} \right) \Big]
\end{array}
\label{eq:interpolatingC}
\end{equation}

Here $2^{3/8}s^{1/4}$ is the limit of $F_ -(s)$ as $s \rightarrow \infty$, which is the scaling part of the exact magnetization $M(T)^2 = (1-\sinh(2/T)^{-4})^{1/4} \approx r^{1/4} F_- (\infty) \approx m^2_{t} t^{1/4}$. Also, we have $m_0 = m_t  |1-\theta_c^2|^{1/8} /\theta_c$. The exact zero-field scaling solutions $F_+$ and $F_-$ are integrals of Painlev{\'e} transcendents of the third kind, which are not expressible in closed form functions of $s$, nor readily available in subroutine libraries. In the supplemental material, we offer a high-precision implementation for the necessary Painlev{\'e} function. In addition, we provide a simple fitting form for $F_+$ and $F_-$, accurate to within 3.4\% and $1\%$ respectively for three orders of magnitude of scaling arguments and written in terms of simple elementary functions.

The interpolating function $f(\theta)$ is chosen to match the scaling form for the susceptibility, namely $\chi(R, \theta) = R^{-7/4} X(\theta) = \int dr 2 \pi r \left( C(r, R, \theta) - m(R, \theta)^2 \right)$. Using a high-precision polynomial form for $X(\theta)$ from~\cite{caselle2001critical} (see supplemental material), this leads to: 
\begin{equation}
f(\theta) = \frac{X(\theta) / 2 \pi - \Xi(\theta)^{7/4} {\cal I}_+}{\Xi(\theta)^{7/4} ({\cal I}_- - {\cal I}_+)}
\end{equation}  
where ${\cal I}_+ = \int dy y^{3/4} F_+(y)$, ${\cal I}_- =  \int dy y^{3/4} (F_-(y) -m_0^2 y^{1/4} \theta_c^2 \Xi(\theta_c)^{1/4})$. Numerical integration of the exact scaling results gives  ${\cal I}_+ = 0.413135114$ and ${\cal I}_- =  0.010959562$. If we require $f(0)=0$ and $f(\theta_0)=1$, these two constraints give ${\cal I}_+ = 0.413134$ and ${\cal I}_- = 0.0104234$ at our current R.

\begin{figure} [htbp]
\includegraphics[scale=0.35]{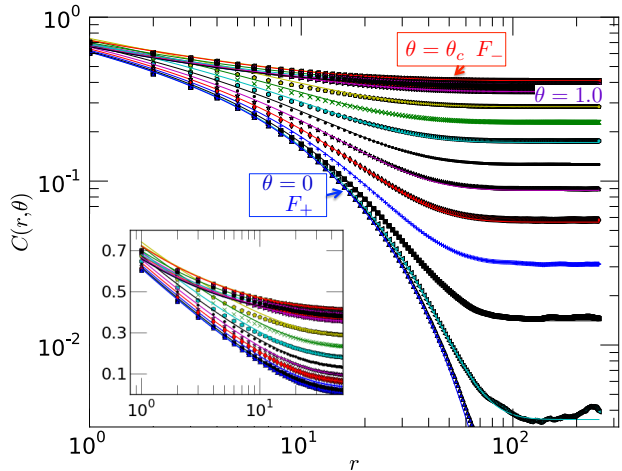}
	\caption{\label{fig:interpolation} Our functional form (lines) for the disconnected correlation function is in excellent agreement with large $r$ simulation data (symbols) from systems of size $\mathrm L = 1024$. Results are plotted at fixed $R$ over the allowed range of $\theta$ as shown in Figure~\ref{fig:ConstantR}.  At small distances (inset), there are other effects that the scaling function fails to capture.}
\end{figure}

% Talk about analytic corrections to scaling now...
% rewrite this section -- how to introduce corrections.
This simple interpolation description for the correlation function (Figure~\ref{fig:interpolation}) is in agreement with simulation results within $2\%$ relative error except at distances less than three spin spacings, below which nonuniversal lattice effects dominate (see supplemental material for discussion). Although our interpolating form is approximate even in the scaling limit, its agreement with simulation results can be improved by including analytic and singular corrections to scaling that vanish with increasing $r$.  

There has been theoretical work~\cite{Aharony1980Corrections,BarmaFisherPRB} on the amplitudes of the dominant analytic corrections to scaling in the 2D Ising model on a square lattice, and evidence that the dominant singular corrections to scaling happen to vanish. In an experimental system (e.g., biomembranes), the magnitudes of these analytic and singular corrections to scaling must be experimentally determined. They will give small corrections near $(T_c, H_c)$, but will extend the validity of the theory further into the phase diagram -- perhaps facilitating systematic identification of phase boundaries.

\begin{figure}[htbp]

\subfigure[$H=0$ $T>T_c$]{\includegraphics[scale=0.3]{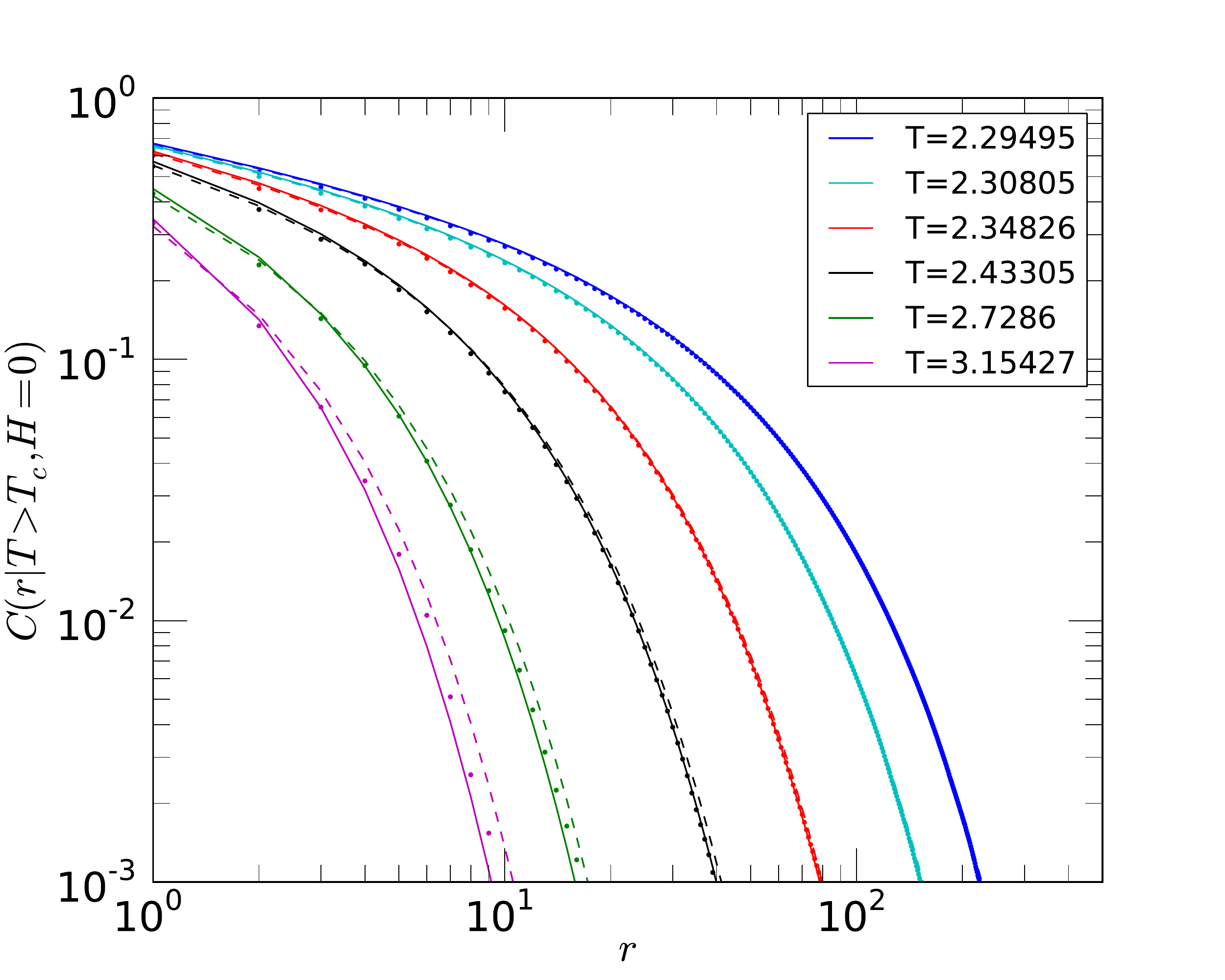}
\label{fig:AnalyTPlus}
}

\subfigure[$H=0$ $T<T_c$]{\includegraphics[scale=0.3]{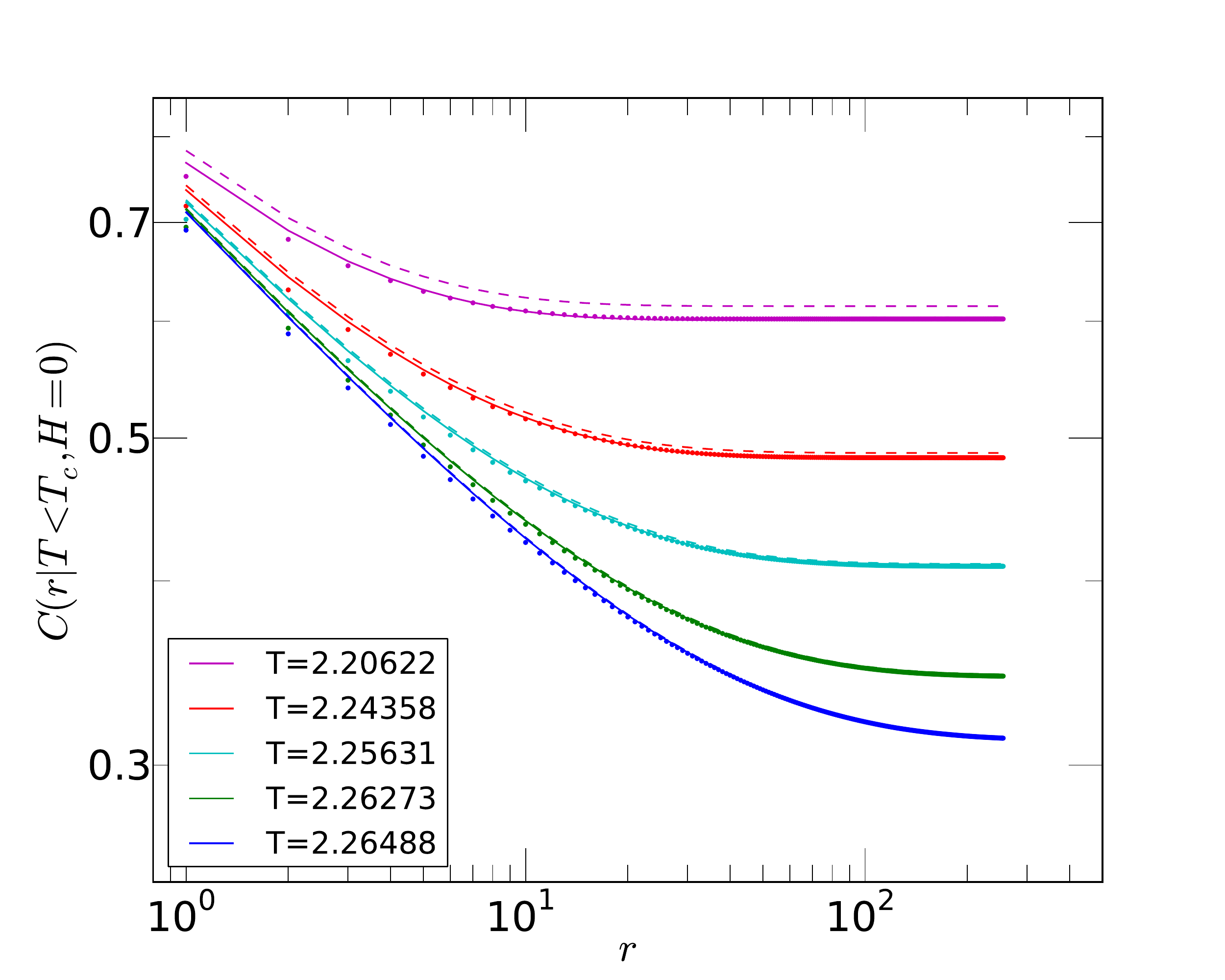}
\label{fig:AnalyTMinus}
}

\caption{{\bf Analytic Corrections} Dots are simulation data from $L=1024$ size simulations using the Wolff algorithm. The dashed line for all the plots are the scaling solutions, with $a(T) \approx a_0$, $\xi(T) \approx \xi_0 t^{-\nu}$, and $u_t = t$.   The solid line is including corrections to first order in $a(T)$, $\xi(T)$, and $u_t$.}
\label{fig:AnalyCorrections}
\end{figure}

We begin at $H=0$, where due to exact results~\cite{wu1976spin} and perturbative studies~\cite{Aharony1983Corrections,BarmaFisherPRB} we can write a general form for both analytic and singular corrections to scaling and can add the leading corrections exactly.  Although Equation~\ref{eq:scalingCorr} becomes exact as $r \rightarrow \infty$, for finite $r$ there are a hierarchy of corrections that arise from a more complete form:
\begin{equation}
C(r|T) = a(T) r^{-\eta} {\cal C} (r/\xi(T), u_3 t^\Delta, \cdots).
\end{equation}
where $u_i$ for $i>2$ are irrelevant directions under the renormalization group, with $u_3$ the leading singular correction. The T-dependent functions can be written in a series expansion of $t$,
\begin{align}
a(T) &= a_0(1+a_{1,0} t + O(t^2)) \\ 
\xi(T) &= \xi_0 t^{-\nu} (1+ c_\xi t + O(t^2)),
\label{eq:correctionsT}
\end{align}
where $a_0 = 2^{3/8}$ and $\xi_0$ = $T_c/4$ are Ising specific prefactors for the scaling form~\cite{wu1976spin}. Expanding the exact form of $a(T)$ and $\xi(T)$ given by McCoy and Wu~\cite{wu1976spin}, we get $a_{1,0} = 2^{-3/2} (4/T_c)$, and $c_\xi = -1/(\sqrt{2} T_c) $ which are the first analytic corrections to scaling. Figure~\ref{fig:AnalyCorrections} shows the improvement in accuracy enabled by these analytic corrections along $H=0$.

%XXX: talk about how experiments might see u_3 and also talk about how these will be hard to separate from analytic corrections.
The first order effect of the leading irrelevant direction $u_3$ to scaling is to generate a power-law, singular correction that looks like $r^{-\eta -\Delta}$:
\begin{align}
&C(r | t, u_3) = r^{-\eta} {\cal C} (r/t^{-\nu}, u_3 t^\Delta) \\
&= r^{-\eta} {\cal C}(r/t^{-\nu}, 0) + u_3(t) r^{-\eta -\Delta} {\cal C}^{(1)} (r/t^{-\nu}, 0) \nonumber
\end{align} 
where ${\cal C}^{(1)} (r/t^{-\nu},  0)$ is the derivative of the scaling function  ${\cal C} (r/t^{-\nu},  u_3 t^\Delta)$ with respect to $u_3$ at $u_3 = 0$.  For the Ising critical point, studies have found that $u_3=0$~\cite{Aharony1983Corrections,BarmaFisherPRB}, and in our data we also see no evidence for a power-law of $r^{-\eta -\Delta}$ upon subtracting the exact scaling solutions from the numerical data.  However, there is no reason that biomembrane experiments should expect $u_3=0$. For example, it is non-zero in Ising-like models (square-lattice Klauder and double-Gaussian model)~\cite{BarmaFisherPRB}, with $\Delta \approx 1.35$ 

One can also expand the non-universal amplitude $a(T)$, noticing $a_{10} t = 2^{3/2} s/r$, to get: $C(r) = r^{-1/4} F_{\pm} (s) \pm 2^{-3/2} r^{-5/4} s F_{\pm} (s) + (9/64) r^{-9/4} s^2 F_{\pm} (s)$.  Estimated values of $\Delta$ give $\eta+\Delta \approx 1.6$ and so in principle:
\begin{align}
&C(r) = r^{-0.25} F_{\pm} (s) \pm 2^{-3/2} r^{-1.25} s F_{\pm} (s) \cr
 &+ \frac{9}{64} r^{-2.25} s^2 F_{\pm} (s) + u_3(t) r^{-1.6} {\cal C}^{(1)} (r/t^{-\nu}, 0).
\end{align}
In the case of Ising-like criticality in experiments, where no exact results are known, the combination of the effect of powers of $r^{-1.25}$, $r^{-1.6}$, and $r^{-2.25}$ would not be easy to disentangle, requiring a fit to a function that looks like $C(r) =  r^{-0.25} {\cal C} (s) + r^{-1.25} {\cal C}_1(s) +  r^{-2.25} {\cal C}_2(s) + c_{u_3} r^{-1.6} {\cal C}^{(1)} (s, 0)$. This fit will likely be sloppy~\cite{waterfall2006sloppy}, with the individual coefficients ill-determined.

Now that we have investigated corrections to scaling at $H=0$, let us return to our main goal. The correlation function that we provide in this paper is a function of both temperature and field, so we will need to consider analytic corrections in a field. In principle, instead of the scaling form correlation function written in Equation~\ref{eq:scalingCorr}, the full correlation function should look like:
\begin{equation}
C(r|T, H) = a(T, H) r^{-\eta} {\cal C} (r/u_t^{-\nu}, u_h/u_t^{\beta \delta}, \cdots).
\end{equation}
Here we incorporate analytic corrections to scaling (as in eqns~\ref{eq:correctionsT}) via the scaling fields $u_t$ and $u_h$, which are only consistent with the tuning parameters $t$ and $h$ up to first order.  We may instead write the non-universal amplitudes and scaling fields in terms of $t$ and $h$. Using the notation of~\cite{Aharony1983Corrections} they are,
\begin{align}
a(T, H) &= a_0(1+a_{1,0} t + a_{0, 2} h^2 + a_{2, 0} t^2 +....) \\
u_t &= t (1+c_t t + O(t^2, h^2)) \\
u_h &= h (1+ c_h t + O (t^2, h^2)) \\
\xi(u_t) &= \xi_0 t^{-\nu} (1+c_\xi u_t + O(u_t^2, u_h^2))
\end{align}
By the inversion symmetry of the Ising lattice, only terms of $h^{2n}$ are should be allowed. As $h$ has the same scaling dimensions as $t^{ \beta \delta} = t^{15/8}$, the leading analytic corrections should be controlled by temperature up to $t^3$ and we only consider leading order corrections arising from $t$.

For the 2D Ising model it is known that $c_t = c_h = \beta_c/\sqrt{2}$~\cite{caselle2001critical}, so we can add all forms of analytic corrections to our scaling function.  For the nonlinear scaling fields $u_t$ and $u_h$, we can translate the effective temperature and field $t_{\text{eff}} = t(1+c_t t)$ and $h_{\text{eff}} = h(1 + c_h t)$ to effective Schofield coordinates, $R_{\text{eff}} = R( 1+ g_1(\theta) R)$ and $\theta_{\text{eff}} = \theta (1+ g_2(\theta) R)$ assuming that since $R$ scales with $t$, they will have the same leading order effects, and assuming an arbitrary form for the corrections dependent on $\theta$.  Using our parametric definition (see eq~\ref{eq:htheta}) we can arrive at closed form expressions for $g_1(\theta)$ and $g_2(\theta)$ (see supplemental material), and incorporate $\theta_{\text{eff}}$ and $R_{\text{eff}}$ into Equation~\ref{eq:interpolatingC}. For the amplitude corrections and scaling function corrections, we include them respectively in $F_\pm(s)$ according to the McCoy/Wu expansion. Including the analytic corrections in $a(T)$, $\xi(T)$, $u_R$ and $u_\theta$ for the constant-$R$ data in Figure~\ref{fig:interpolation} leads to a improvement in the accuracy for the data sets at $\theta=0$, $\theta=1$, and $\theta=\theta_c$ (Figure~\ref{fig:AnalyCorrections}), but no systematic improvement for the data sets in between these special points.  This is perhaps unsurprising so near to the critical point, where the analytic corrections are small compared to the residual errors in our scaling form. Overall, including the corrections to scaling, our functional form has an average relative error of $1.5\%$ per data point.

We have constructed a functional form for the spin-spin correlation function of the 2D Ising model at arbitrary temperature and field, which matches known analytical results at $H=0$, and numerical simulations to high precision. By virtue of the real-space representation, comparing Ising predictions to laboratory experiments is reduced to the relatively simple matter of converting experimental parameters (such as biomembrane temperature and composition) to the Ising variables of temperature and magnetization or magnetic field or, equivalently, $R$ and $\theta$. This makes our functional form a robust tool for testing whether or not real systems fall into the Ising universality class.

We would like to thank Lorien Hayden for help in debugging the Ising simulations, Alex Alemi for recommending the Chebyshev polynomials, and Michael Fisher for informative correspondence. YJC and JPS acknowledge support from NSF DMR-1005479 and DMR-1312160.  NMP is supported by a Stanford Humanities and Sciences Fellowship and was partially supported by a Cornell Rawlings Presidential Research Scholarship.  BBM was supported by NIH T32GM008267 and a Lewis-Sigler Fellowship.

%\bibliographystyle{apsrev}
%\bibliography{IsingPaperRefs.bib}

\clearpage

%\title{Supplemental Material: 2D Ising Model: Precise functional form for the spin-spin correlation function}

%\maketitle
\begin{center}
{\large {\bf Supplemental Material}}
\end{center}

\section{Scaling solutions at $H=0$}
\subsection{Numerical evaluation of scaling solutions}
In the main text, we make use of the scaling solutions for the disconnected correlation function at $H=0$,
\begin{equation}
\langle \sigma_r \sigma_0 \rangle = r^{-1/4} F_{\pm} (s)
\end{equation}
which is valid as $T \rightarrow T_c$ and $r \rightarrow \infty$ with $s$ fixed. The symbol $\pm$ denotes solutions for $T > T_c$ and $T< T_c$ respectively.  Analytical studies use $s^* = |z^2 +2 z -1|/\sqrt{z (1-z^2)} r$ with $z = \tanh(1/T)$ as the argument for this function, however in the main text we used the scaling form $s = (4/T_c) (r/t^{-\nu})$.   The solutions are of the form~\cite{wu1976spinb}:
\begin{widetext}
\begin{equation}
F_{\pm}(s) =  2^{-1/2} (2 \sinh(2/T))^{1/8} (s/2)^{1/4} \left( 1\mp \eta(s/2) \right) \eta(s/2)^{-1/2} \exp \left( \int_{s/2}^\infty dx \frac{x}{4} \eta(x)^{-2} ((1-\eta(x)^2)^2 - \eta^\prime (x)^2) \right)
\label{eq:FPlusMinus}
\end{equation}
\end{widetext}
$\eta(\theta)$ is the solution to the Painlev{\'e} differential equation of the third kind,
\begin{equation}
\frac{d^2 \eta}{d \theta^2} = \frac{1}{\eta} \left( \frac{d \eta}{d \theta} \right)^2 - \eta^{-1} + \eta^3 - \theta^{-1} \frac{d \eta}{d \theta}
\end{equation}
with boundary conditions
\begin{equation}
\eta(\theta) = -\theta \left[ ln\left(\frac{\theta}{4}\right) + \gamma_E \right] + O(\theta^5 ln^3 \theta)
\label{eq:smalleta}
\end{equation}
as $\theta \rightarrow 0$, and
\begin{equation}
\eta(\theta) = 1- \frac{2}{\pi} K_0(2 \theta) + O(e^{-4 \theta})
\label{eq:largeeta}
\end{equation}
as $\theta \rightarrow \infty$ with $K_0(x)$ is a modified Bessel function of the 2nd kind.
The Painleve transcendent $\eta(\theta)$ is not expressible in terms of elementary functions; to evaluate it numerically, we choose to use tools available in the Chebfun Matlab package~\cite{chebfunv4}. We use a Chebyshev polynomial approximation for $\eta(\theta)$ between arguments of 0.003 and 3, while we use the asymptotics given in Equations~\ref{eq:smalleta} and~\ref{eq:largeeta} for arguments outside of this range. To evaluate Equation~\ref{eq:FPlusMinus}, we use integration subroutines available in Matlab and Python, the adaptive Simpson quadrature function \textit{quad} in Matlab, and the scipy.integrate.quad function which draws from the Fortran library QUADPACK (mainly adaptive quadrature techniques).   Our Matlab implementation and the Python module containing the Chebyshev polynomial for $\eta(\theta)$ and the $F_{\pm} (s)$ scaling function are available online~\cite{supplementalWeb}.

\subsection{Effective Functional Form}

For convenience and less opaque representation of the scaling solutions, we also provide an effective functional form, which is good to $3.4\%$ relative accuracy for $F_+$ and $1\%$ relative accuracy for $F_-$. These functions are an interpolation between the small and large distance asymptotics for the exact scaling solutions at $H=0$.

Both scaling functions have $F_\pm (0) = C_0 = 0.7033801577...$. The asymptotic large-r behavior is different depending on whether one is above or below criticality. The $T>T_{c}$ case is particularly simple, partially since $\langle M^{2} \rangle = 0$. We simply choose the effective large-r functional form to be the exponential decay given by the Ornstein-Zernike decay, which is like $s^{-1/4} \exp(-s)$ for $T>T_c$. The amplitude of this piece, called $p_1$, is determined by an asymptotic expansion of the large distance Bessel functions, and we get $p_1 = 1/(2^{1/8} \sqrt{\pi})$.

We find a simple and effective nonlinear interpolation that we will employ in both the high and low-temperature cases. Empirically, we find that both functions are well-described by $F^{\text{fit}}_{\pm} = (B(s)^{|k|} (\text{Small-r})^{k} + (1-B(s)^{|k|}) (\text{Large-r})^{k})^{1/k}$, where $k$ is a fit parameter, that controls the nonlinear interpolation of the functions, whereas a weighting function $B(s)$ that has the limits $B(0)=1$ and $B(\infty) = 0$ controls the weight of each piece of the interpolation. For $T>T_c$ we write:
\begin{equation}
F^{\text{fit}}_{+}(s)  = \left(0.70338^{k} B(s)^{|k|} + (1-B(s))^{|k|} (p_1 \cdot s^{-1/4} \exp(- s))^{k} \right)^{1/k}.
\label{eq:Fplus}
\end{equation}
with
\begin{equation}
B(s) = \exp(-(c x)^b)
\end{equation}
If $k$ is negative, we need to make the weights $(1/B(s))^{k}$ and $(1/(1-B(s)))^{k}$, for $F^{\text{fit}}_+(s)$ to have the right limits at $s=0$ and $s=\infty$, hence the absolute value $|k|$ in the power of those terms.

\begin{figure}[htbp]
\begin{center}
\includegraphics[scale=0.35]{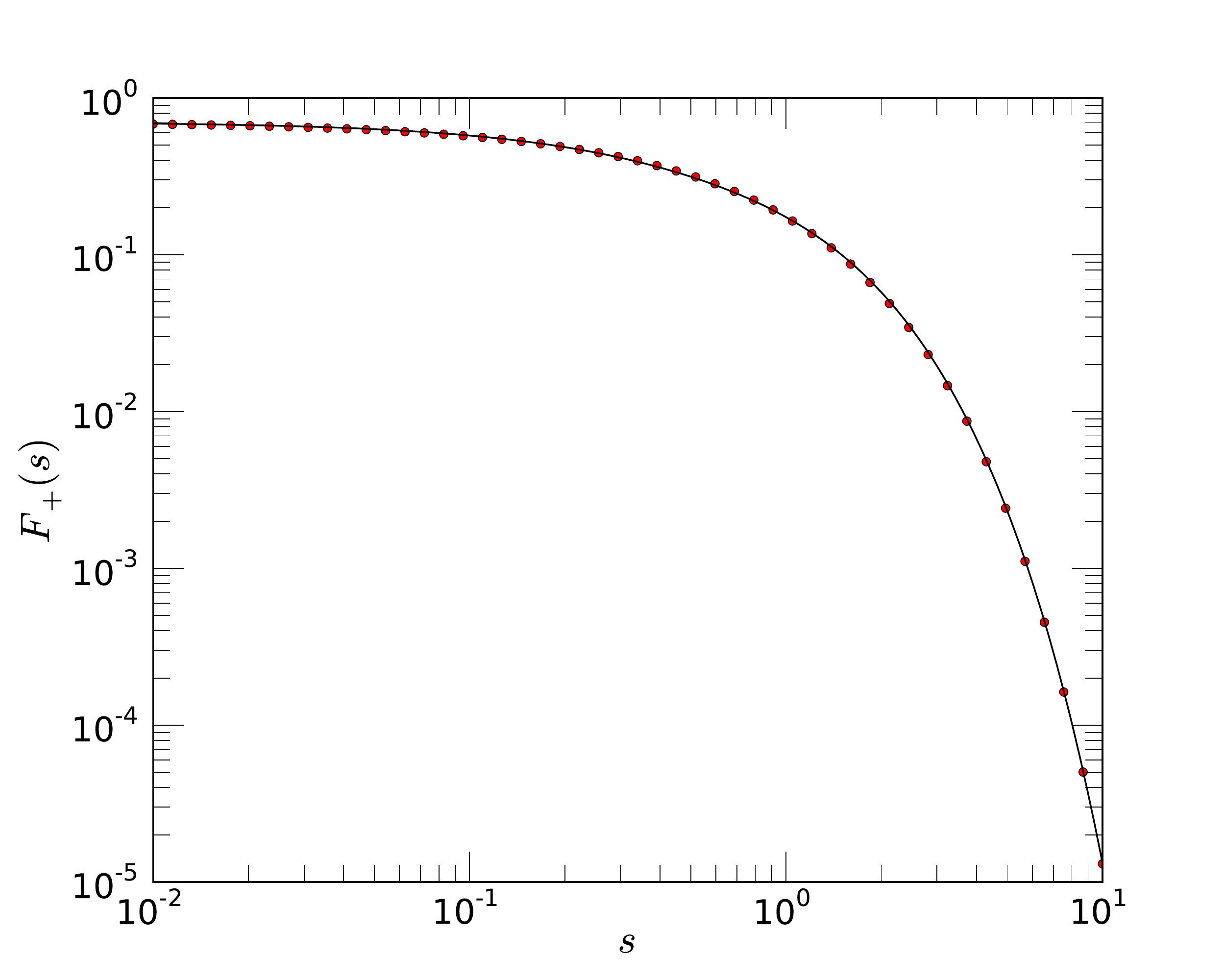}
\caption{{\bf fit to Painleve results} This is a fit to equation~\ref{eq:Fplus}. Fit parameters $c = 1.73$, $b =0.92$, $k=3.8$.  Red dots are Painleve results, and the black line is the result of the fit.}
\label{fplusFit}
\end{center}
\end{figure}

Our form matches the exact solution to within $3.4\%$ maximum relative error, with an average of $1.5\%$ error, for the range of our fit $0.01 \le s \le 10$. (See figure \ref{fplusFit}).

Now let's turn our attention to the $T<T_c$ case. The philosophy for constructing the effective functional form is identical to the high temperature case, although for the disconnected correlation function, the long distance asymptote is dominated by the magnetization $\langle M \rangle^2$. For the connected correlation function $\langle \sigma_0 \sigma_r \rangle - \langle \sigma_0 \rangle \langle \sigma_0 \rangle$, with the magnetization squared subtracted off, the long distance decay for the scaling function is $p_2 s^{-7/4} \exp(-2 s)$, with $p_2 = 1/(2^{21/8} \pi) $.  In our effective functional form, for simplicity, we choose to fit only the connected correlation function, interpolating between the short distance behavior and long distance decay, while add the scaling magnetization squared to the result. (if one wishes, analytic corrections may be incorporated into the scaling magnetization as well). We use:
\begin{equation}
\begin{split}
F^{\text fit}_{-} (s) =& \left((B(s) \cdot 0.700883)^k + ((1-B(s))(p_2 s^{-7/4} \right. \\
&\left.  \vphantom{(B(s))^k}\exp(-s)))^k \right)^{1/k} + 2^{3/8} s^{1/4}
\label{eq:nomagFminus}
\end{split}
\end{equation}
Here,
\begin{equation}
B(s) = \exp(-(s/c)^b)
\end{equation}
From fits, we find $c=0.007 \pm 0.07$, $b= 0.4 \pm 2$, and $k=-0.2\pm0.1$ .
\begin{figure}[htbp]
\begin{center}
\includegraphics[scale=0.35]{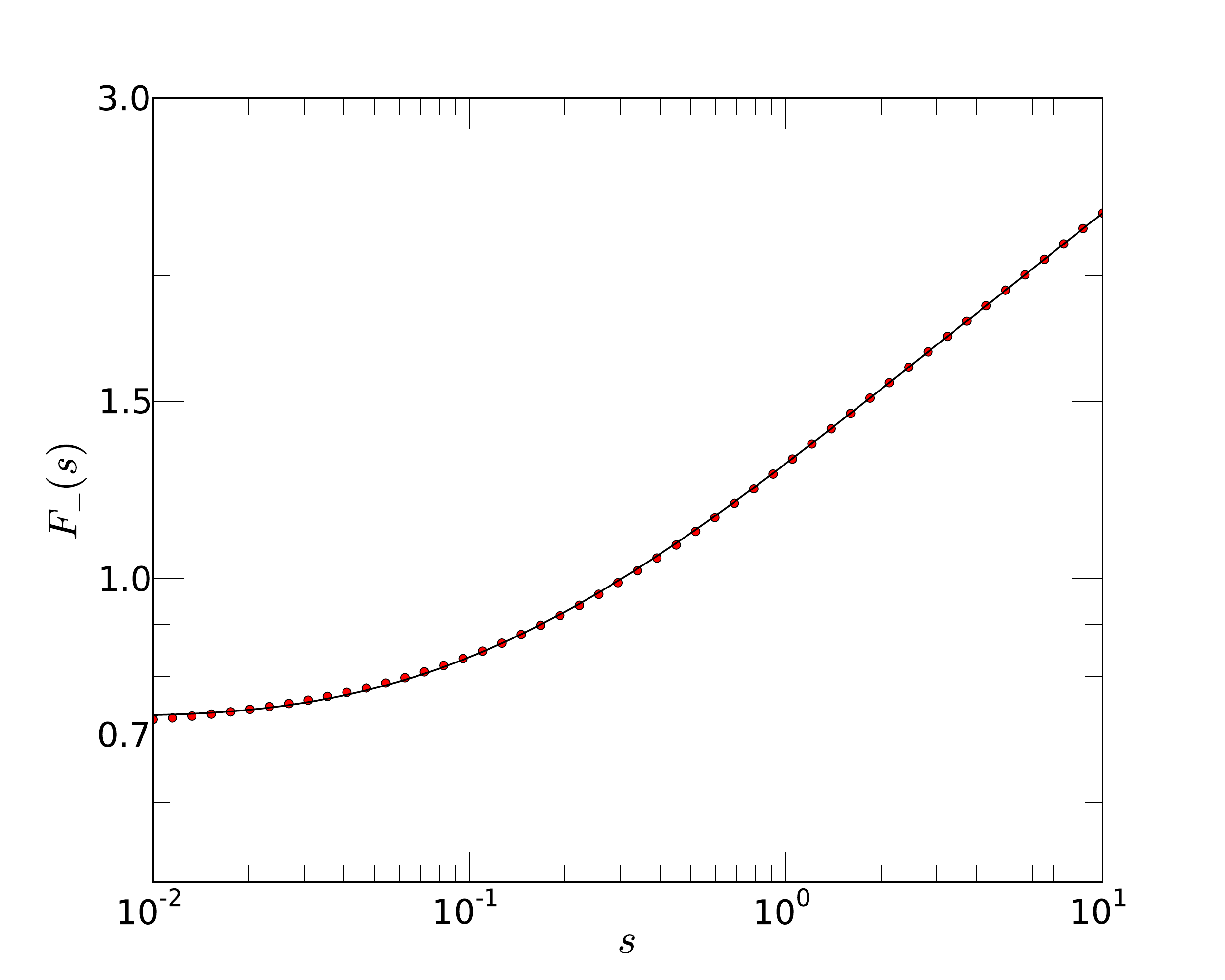}
\caption{{\bf fit to Painleve results} This is a fit to equation~\ref{eq:nomagFminus}. Fit parameters are: $c=3.62$, $b=0.87$, and $k=-0.4$.  The magnetization is separated out, so that we may subtract it off for the interpolation- which makes things less complicated. Red dots are Painleve results, and the black line is the result of the fit.}
\label{fminusFitTwo}
\end{center}
\end{figure}
The fit is good to a maximum of $1\%$ error when compared against our Chebyshev form for the range of our fit $0.01 \le s \le 10$ (Figure~\ref{fminusFitTwo}).

\section{High-precision scaling form for the susceptibility}
We use a high-precision form of the susceptibility as an integration constraint for our functional form. The susceptibility was derived from the high-precision approximate forms for the equation of state in Reference~\cite{caselle2001criticalb}. Using the parametric representation
\begin{equation}
\begin{split}
t &= \frac{T-T_c}{T} = R(1-\theta^{2}) \\
h &= H/T = h_{0} R^{\beta\delta}h(\theta) \\
M &= m_{0} R^{\beta} \theta
\end{split}
\end{equation}
the high-precision form for the equation of state for $h(\theta)$ is
\begin{equation} 
\begin{split}
h(\theta) =& \left( \theta - \frac{\theta^3}{1.16951} \right) (1 - 0.222389 \theta^2 - 0.043547\theta^4 \\
&- 0.014809\theta^6 - 0.007168\theta^8),
\end{split}
% \label{eq:htheta}
\end{equation}
and the definition $\chi = d M / d h$, we have:
\begin{widetext}
\begin{equation}
\chi(R, \theta)= R^{-7/4}m_0 \frac{\left(1+(2\beta-1) \theta ^2\right)}{h_0 \left(1-0.482344 \theta ^2-0.0750424 \theta ^4-0.0262771 \theta ^6-0.0234342 \theta ^8+0.0385732 \theta ^{10}-0.0444357 \theta ^{12}\right)}
\end{equation}
\end{widetext}

\section{Analytic Corrections to Scaling}
The analytic corrections to scaling to the RG field $u_t$ and $u_h$ are given in coordinates of $t$ and $h$ in the literature. Since we give our function in parametric coordinates, here we provide forms for the corrections to be expressed in $R$ and $\theta$. In the main text we state that:
\begin{align}
u_t &= t (1+ c_t t + O(t^2)) \\
u_h &= h(t+c_h t + O(t^2)).
\end{align}
Since $t = R (1-\theta^2)$, $R$ scales with $t$, so the first order corrections should also be linear in $R$.  However, $\theta$ is not small as it can take any value from $0$ to $\theta_c \approx 1.08144...$, so we will assume that 
\begin{align}
u_R &= R (1 + g_1(\theta) R + O(R^2)) \\
u_\theta &= \theta (1+ g_2(\theta) R + O(R^2)).
\end{align}
We can then solve for $g_1(\theta)$ and $g_2(\theta)$ using the Schofield definition:
\begin{widetext}
\begin{align}
g_1(\theta) &= \frac{ c_h \theta^3 \left(2-6.1549 \theta^2+6.60301 \theta^4-2.69648 \theta^6+0.214502 \theta^8+0.0351324 \theta^{10}-0.0135271 \theta^{12}+0.0122581 \theta^{14}\right) }{\left(\theta -1.48234 \theta ^3+0.407301 \theta ^5+0.0487653 \theta ^7+0.00284291 \theta ^9+0.0620074 \theta ^{11}-0.0830089 \theta ^{13}+0.0444357 \theta^{15}\right)} \cr
&+ \frac{c_t \theta  \left(1-6.23234 \theta ^2+13.4301 \theta ^4-12.7392 \theta ^6+5.00997 \theta ^8-0.343026 \theta ^{10}-0.210889 \theta ^{12}+0.152808 \theta ^{14}-0.0674197 \theta ^{16}\right)}{\left(\theta -1.48234 \theta ^3+0.407301 \theta ^5+0.0487653 \theta ^7+0.00284291 \theta ^9+0.0620074 \theta ^{11}-0.0830089 \theta ^{13}+0.0444357 \theta^{15}\right)} \\
g_2(\theta) & = \frac{\left((c_h - \beta \delta c_t) \left(1-\theta^2\right)^2 \left(\theta -0.855059 \theta ^3\right) \left(1.-0.222389 \theta ^2-0.043547 \theta ^4-0.014809 \theta ^6-0.007168 \theta ^8\right)\right)}{\left(\theta -0.482344 \theta ^3-0.0750424 \theta ^5-0.0262771 \theta ^7-0.0234342 \theta ^9+0.0385732 \theta^{11}-0.0444357 \theta^{13}\right)}
\end{align}
\end{widetext}

\section{Accuracies and Errors}

Here we report the quality our interpolation form in terms of average cost per data point, and average relative error per data point for each of the simulation datasets at R = 0.0336737 with and without analytic corrections. We define the un-weighted residual to be: 
\begin{equation}
r_j = D(j, \theta, R) - C(j, \theta, R)
\end{equation}
where $D(j, \theta, R)$ is the data, $C(j, \theta, R)$ the interpolating form. The average cost was calculated with the covariance matrix multiplying the residual:
\begin{equation}
{\text cost} = r_i \sigma^{\text cov}_{ij} r_j / N.
\label{eq:avcost}
\end{equation}  The relative error was measured as $\langle e_{\text rel}^2 \rangle$,
where
\begin{equation}
e_{\text rel} = \frac{D(j, \theta, R) - C(j, \theta, R)}{C(j, \theta, R)} . 
\label{eq:relerror}
\end{equation}  

\begin{table}[htdp]
\begin{center}
\begin{tabular}{c|c|c|c|c}
$\theta$ & h& T& cost & error (\%) \cr
\hline
0 &0&2.348260&0.5&2.85 \cr
0.10&1.612125e-04&2.347442&5.1&1.74 \cr
0.20&3.119619e-04&2.344991&2.9&3.55 \cr
0.30&4.420782e-04&2.340918&3.4&1.45 \cr
0.40&5.419973e-04&2.335240&3.8&2.72 \cr
0.50&6.031200e-04&2.327979&4.5&2.14 \cr
0.60&6.182554e-04&2.319166&5.1&0.91 \cr
0.70&5.822216e-04&2.308836&6.1&2.17 \cr
0.80&4.927350e-04&2.297031&8.7&1.96 \cr
0.90&3.518271e-04&2.283797&8.6&1.28 \cr
1.00&1.681982e-04&2.269185&0.7&0.20 \cr
1.01&1.481081e-04&2.267650&2.1&0.53 \cr
1.02&1.277970e-04&2.266102&2.4&2.58 \cr
1.03&1.072934e-04&2.264541&6.3&0.86 \cr
1.04&8.662751e-05&2.262967&6.8&1.00 \cr
1.05&6.583165e-05&2.261380&6.9&1.02 \cr
1.06&4.494000e-05&2.259780&5.5&0.92 \cr
1.07&2.398890e-05&2.258167&3.1&0.95 \cr
1.08&3.016912e-06&2.256541&1.0&0.19 \cr
$\theta_c$&0&2.256306&3.6&0.85 \cr
\end{tabular}
\end{center}
\caption{{\bf Cost and Errors for Interpolation} The quality of our interpolation function is tabulated here in terms of average relative error (Equation~\ref{eq:relerror}) and average cost (Equation~\ref{eq:avcost}). For the calculation of this table, we skip the first 3 points (where lattice effects and higher-order corrections to scaling dominate) and data for $C<10^{-2}$ (where the error is dominated by insufficient numerical statistics). (Note that the only data sets with values smaller than $10^{-2}$ are $\theta=0$ and $\theta=0.1$.) For $\theta=0$, the statistical error becomes comparable to the data value once $C(r)<0.01$, the error approaches $50 \%$ of the data value and exceeds that once $C(r)<0.01$, and for $\theta=0.1$ it approaches $5-10\%$ after $C(r)<0.01$. We expect our scaling form to be excellent in these large-distance regimes, where the corrections to scaling are negligible and the effects of the external field are small.}
\label{table:accuracies}
\end{table}%

\begin{table}[htdp]
\begin{center}
\begin{tabular}{c|c|c|c}
$\theta_{\text eff}$ & $R_{\text eff}$ & cost &  error (\%) \cr
\hline
0.0000000&0.0339815&0.8&1.81 \cr
0.0992207&0.0340277&1.4&1.43 \cr
0.1985608&0.0340413&1.1&2.91 \cr
0.2981203&0.0340234&1.2&2.11 \cr
0.3979613&0.0339790&1.4&1.95 \cr
0.4980952&0.0339168&1.3&1.41 \cr
0.5984751&0.0338478&1.3&1.28 \cr
0.6989978&0.0337834&1.6&1.52 \cr
0.7995184&0.0337330&1.5&1.49 \cr
0.8998839&0.0336999&1.4&1.34 \cr
1.0000000&0.0335563&0.1&0.14 \cr
1.0099992&0.0336703&0.5&0.53 \cr
1.0199971&0.0336664&2.0&2.49 \cr
1.0299941&0.0336619&0.9&0.95 \cr
1.0399909&0.0336568&1.0&1.06 \cr
1.0499883&0.0336507&1.1&1.10 \cr
1.0599874&0.0336434&0.9&0.91 \cr
1.0699899&0.0336345&0.7&0.74 \cr
1.0799981&0.0336234&0.6&0.56 \cr
1.0814389&0.0336216&0.4&0.43 \cr
\end{tabular}
\end{center}
\caption{{\bf Cost and Errors with Corrections} Here are the accuracies of the interpolation with all first order corrections (for $a(T)$, $\xi(T)$ $u_t$, and $u_h$) reported in terms of average relative error (Equation~\ref{eq:relerror}) and average cost (Equation~\ref{eq:avcost}). As in Table~\ref{table:accuracies}, we skip the first 3 points, and data below $10^{-2}$. Note that with corrections the errors are smaller.}
\label{table:accuracies_corrections}
\end{table}%

We may note that the relative error and cost do not necessarily reflect the same measure of theory quality. Relative error gives us a measure  of the level of accuracy for the theory numbers, irrespective of how large the error bars are on the data. Cost, on the other hand, is weighted by the error of the data, and when the average cost is near or smaller than 1.0, the error is mainly caused by statistical fluctuations in the data.  The higher the cost, the less well the theory is capturing the data to within error bars.

The accuracies in Table~\ref{table:accuracies} and~\ref{table:accuracies_corrections} were calculated for distances where the value of the disconnected correlation function $C(j,\theta,R) >0.01$, this means skipping the points above $r=44$ for $\theta=0$, and $r=51$ for $\theta=0.1$.  This is due to the fact that for $\theta=0$ and  $r>44$, the errors are around $50\%$ of the data value. We've also skipped the first $3$ points of each data set due to the fact of short distance corrections dominated by lattice effects of higher-order analytic corrections to scaling (see next section).  

Notice in both tables that the special points $\theta=0$ and $\theta=\theta_c$ whose exact results we interpolate between have a cost that is relatively small, and also that the corrections to scaling improves the overall accuracy. As noted in the main text, the analytic corrections to scaling are small, and do not uniformly improve fits away from these special values. This is not surprising. Since the analytic corrections to scaling this close to the critical point are smaller than our interpolation errors in the scaling function, we might expect they would have cancelling effects roughly half the time. The analytic corrections should be of significant value farther from the critical point at all fields and temperatures.

\section{Small Distance Discrepancies}
\label{sec:smalldistance}

\begin{figure}[htbp]
\begin{center}
\includegraphics[scale=0.35]{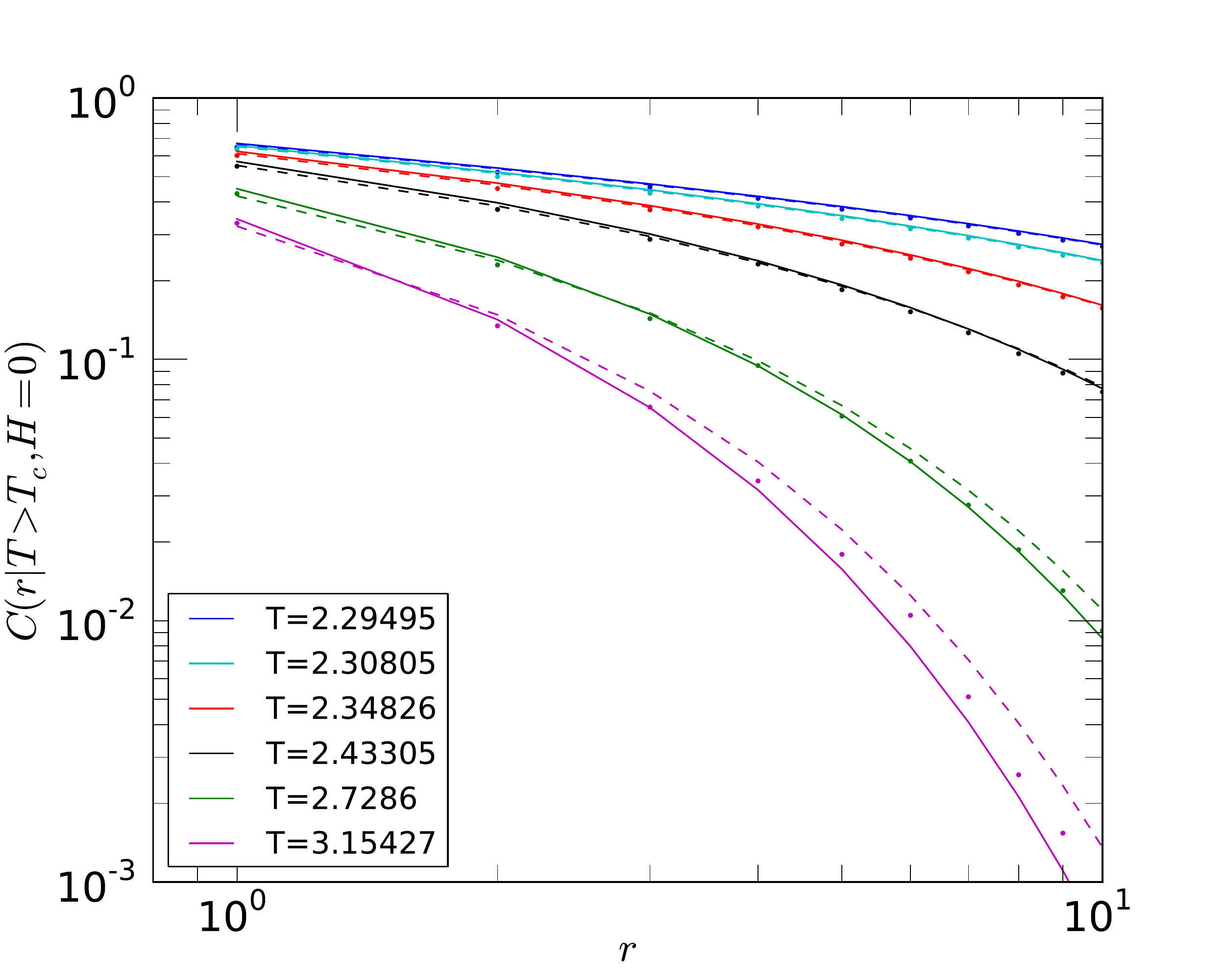}
\caption{\textbf{Small Distance Discrepancies for $T>T_c$ $H=0$} This figure shows the small distance discrepancies for data along $T>T_c$, $H=0$. The dashed line is the scaling theory, while the solid line is including all first order corrections in $a(T)$, $\xi(T)$, and $u_t$. One can see that the discrepancy between simulation data and theory gets smaller as the distance increases.}
\label{fig:smallRTPlus}
\end{center}
\end{figure}

\begin{figure}[htbp]
\begin{center}
\includegraphics[scale=0.35]{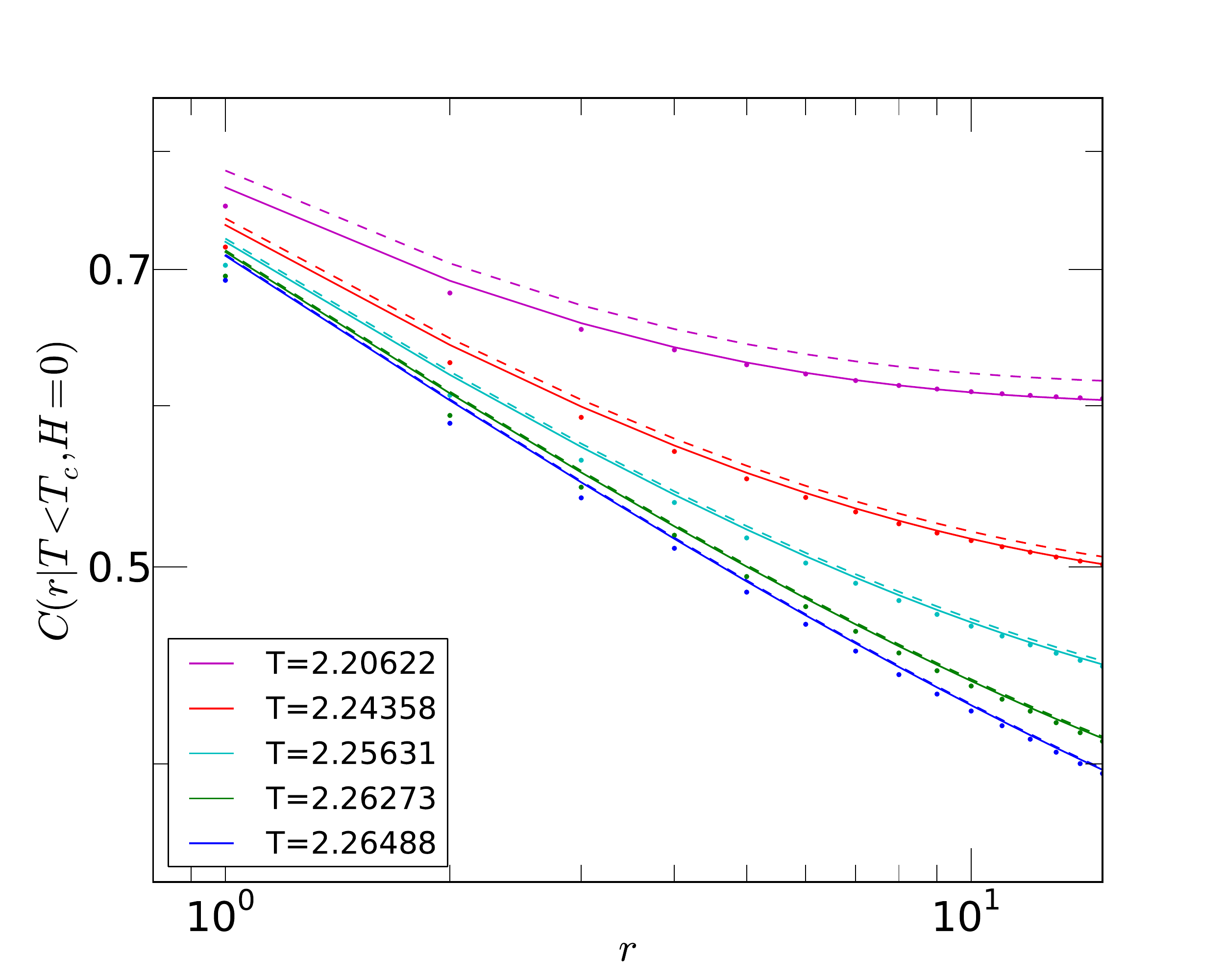}
\caption{\textbf{Small distance corrections for $T<T_c$ $H=0$} This figure shows the small distance discrepancies for data along $T<T_c$, $H=0$. The dashed line is the scaling theory, while the solid line is including all first order corrections in $a(T)$, $\xi(T)$, and $u_t$. One can see that the discrepancy between simulation data and theory gets smaller as the distance increases.}
\label{fig:smallRTMinus}
\end{center}
\end{figure}

\begin{figure}[htbp]
\begin{center}
\includegraphics[scale=0.35]{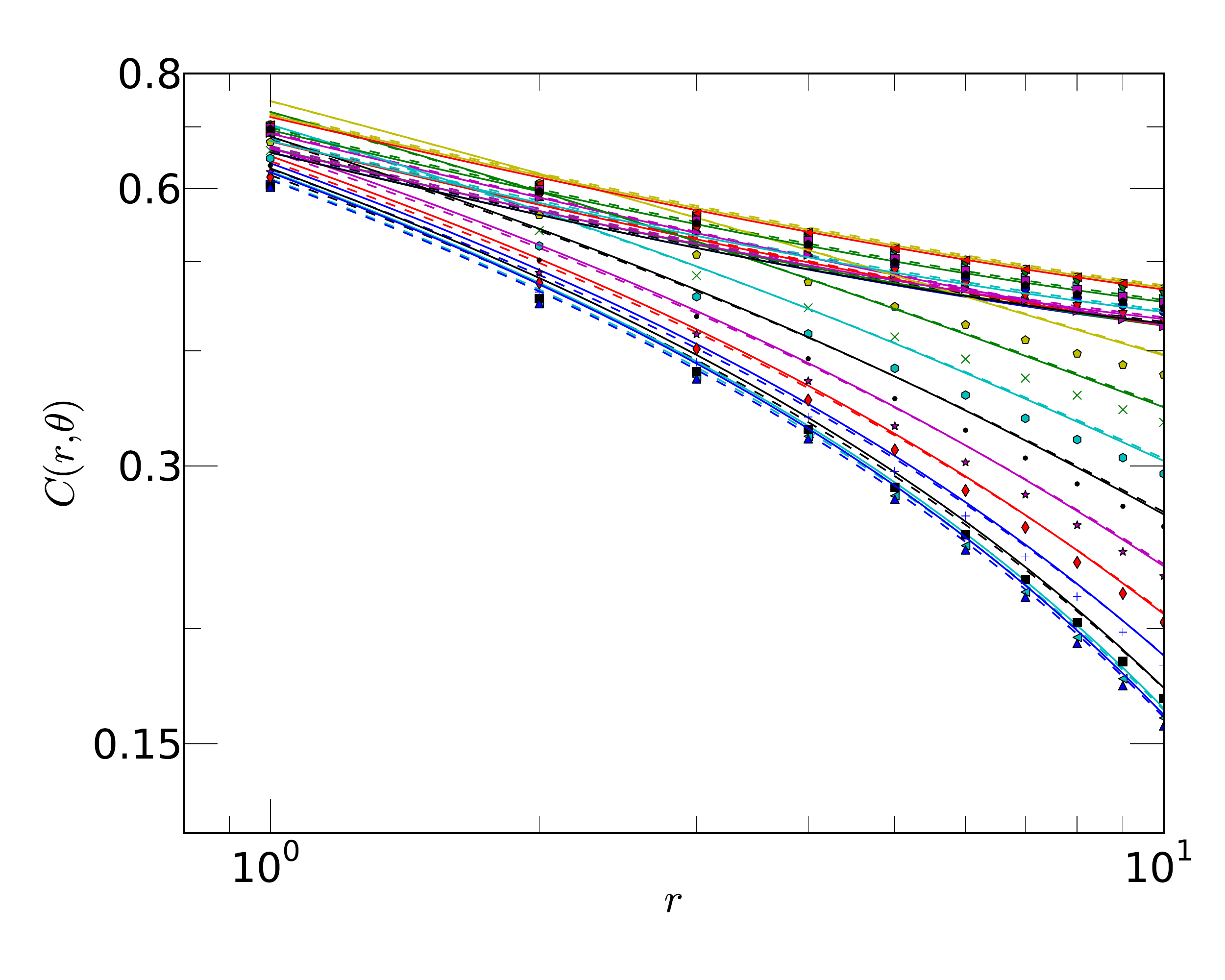}
\caption{\textbf{Small Distance Corrections for Varying $H$ and $T$} This is a plot that shows the small distance discrepancies for all the data we included in matching our interpolation results. It is a larger version of the inset of Figure~2 of the main paper, so that we may see the details of where the scaling theory fails.}
\label{fig:smallRAllTheta}
\end{center}
\end{figure}

The scaling solutions differ from the numerical data at small distances, as shown in Figures~\ref{fig:smallRTPlus},~\ref{fig:smallRTMinus} and~\ref{fig:smallRAllTheta}. We have investigated where this discrepancy stems from, by looking along the $H=0$ axes where exact scaling solutions are known, and consistent with the literature we see no existence of singular corrections (which would be indicated by a power law), nor do we see a dependence between the discrepancy and the distance from the critical point. Most likely the small distance discrepancy is due to the fact that the form of the scaling solution goes as $C_{\mathrm theory}(r) \sim a_0 r^{-1/4}$ for small distances, diverging as $r \rightarrow 0$, however for any data, $C_{\mathrm data}(0) = 1.0$. Therefore, the ratio between the theory and data $C_{\mathrm theory}/C_{\mathrm data}$ diverges as $r \rightarrow 0$.  We attempted to multiply our function by $1/\exp(A/r)$ or equivalently $\exp(-A/r)$ with $A>0$ to incorporate the lattice corrections, but a fit to with this correction does not noticeably improve the quality of our fit.

\section{Numerical Methods}
\subsection{Wolff Algorithm in a field}
The Wolff algorithm~\cite{Wolff89} efficiently simulates the 2D Ising model in zero field, and requires small modifications to be used in non-zero magnetic field.  In the usual Wolff algorithm, which generates members of the ensemble of the Ising model in zero magnetic field, a random spin is chosen which 'seeds' a cluster.  All of the nearest neighbors of this new cluster that have the same spin are then stochastically added to the cluster with the Wolff Probability, $P_{\text{Wolff}}=1-e^{-\beta J}$.  The nearest neighbors of these new additions to the cluster are again added with the Wolf probability, and this process is iterated until a step adds no new spins to the cluster.  At this juncture, the entire cluster is flipped.  To implement a positive magnetic field, $h$, we distinguish between clusters which flip spins from up to down, and those that flip spins from down to up.  Clusters that flip spins from down to up proceed as usual, but whenever an up spin is added to a down cluster, the entire cluster is rejected stochastically with probability $1-\exp(-h)$.  In implementing this algorithm, we were careful to use a predetermined number of proposed cluster flips, rather than a set number of spins, or successful cluster flips.

We implemented all of our simulations on $1024 \times 1024$ square lattices.  Equilibration times were conservatively estimated by waiting for many times the amount of time it takes for the magnetization to reach and then oscillate around its long time value.  After equilibrating we determined the approximate correlation time: the number of proposed clusters that must on average be flipped to generate a new configuration whose magnetization is almost uncorrelated with the previous one.  We generated 100 such independent configurations for each $h$ and $t$ value, and used these to estimate the correlation functions.

\end{document}